\documentclass{elsarticle}
\makeatletter
\def\ps@pprintTitle{%
 \let\@oddhead\@empty
 \let\@evenhead\@empty
 \def\@oddfoot{\textcopyright 2018. This manuscript version is made available under the \href{http://creativecommons.org/licenses/by-nc-nd/4.0/}{CC-BY-NC-ND 4.0 license}.}%
 \let\@evenfoot\@oddfoot}
\makeatother
\usepackage{hyperref, listings, fancyvrb, tikz, mathtools}
\usepackage{comment}
\usepackage{caption, subcaption}
\usetikzlibrary{shapes, arrows}
\usepackage{lineno,hyperref}
\usepackage{amssymb}
\usepackage{textcomp}

\journal{Nuclear Instruments and Methods in Physics Research Section A}

\begin{document}

\begin{frontmatter}

\title{The MARS15-based FermiCORD code system for calculation of the accelerator-induced residual dose}
%\tnotetext[mytitlenote]{Fully documented templates are available in the elsarticle package on %\href{http://www.ctan.org/tex-archive/macros/latex/contrib/elsarticle}{CTAN}.}

%% Group authors per affiliation:
\author{A. Grebe}
\ead{agrebe@mit.edu}
%\address{Washington University in St.~Louis, St.~Louis, MO 63105, USA}
\address{Massachusetts Institute of Technology, Cambridge, MA 02139, USA}
%\fntext[myfootnote]{Since 1880.}

\author{A. Leveling}
%\ead{leveling@fnal.gov}
\address{Fermi National Accelerator Laboratory, Batavia, IL 60510-5011, USA}

%% or include affiliations in footnotes:
\author{T. Lu}
%\ead{tlu@imsa.edu}
%\address{Illinois Mathematics and Science Academy, Aurora, IL 60506-1000, USA}
\address{Carnegie Mellon University, Pittsburgh, PA 15213, USA}

\author{N. Mokhov}
%\ead{mokhov@fnal.gov}
\address{Fermi National Accelerator Laboratory, Batavia, IL 60510-5011, USA}

\author{V. Pronskikh}
\ead{vspron@fnal.gov}
\address{Fermi National Accelerator Laboratory, Batavia, IL 60510-5011, USA}

\begin{abstract}
The FermiCORD code system, a set of codes based on MARS15 that calculates the accelerator-induced residual doses at experimental facilities of arbitrary configurations, has been developed. FermiCORD is written in C++ as an add-on to Fortran-based MARS15. The FermiCORD algorithm consists of two stages: 1) simulation of residual doses on contact with the surfaces surrounding the studied location and of radionuclide inventories in the structures surrounding those locations using MARS15, and 2) simulation of the emission of the nuclear decay $\gamma$-quanta by the residuals in the activated structures and scoring the prompt doses of these $\gamma$-quanta at arbitrary distances from those structures. The FermiCORD code system has been benchmarked against similar algorithms based on other code systems and against experimental data from the CERF facility at CERN, and FermiCORD showed reasonable agreement with these. The code system has been applied for calculation of the residual dose of the target station for the Mu2e experiment and the results have been compared to approximate dosimetric approaches.
\end{abstract}

\begin{keyword}
Monte-Carlo \sep MARS15 \sep residual activation
\MSC[2010] 00-01\sep  99-00
\footnotesize{

Published in \emph{Nuclear Instruments and Methods A} as \href{https://doi.org/10.1016/j.nima.2017.08.055}{doi:10.1016/j.nima.2017.08.055}}
\end{keyword}

\end{frontmatter}

%\linenumbers

\section{Introduction}

High residual radiation doses are an important issue in all accelerator-based experiments, both collider and fixed-target. The particle beams accelerated to relativistic energies either collide with each other or impinge upon targets, producing fluxes of secondary particles.
Among these fluxes, there are substantial fractions of high-energy stable particles arising in spallation and fragmentation reactions, capable of inducing inelastic reactions in the structural materials surrounding those targets and collision points and leading to nuclear transmutations in these structures~\cite{krasa}.

As a result of neutron activation of the walls and other materials that have been exposed to particle fluxes, the structures become radioactive, emitting $\alpha$, $\beta$, and $\gamma$ radiation even after the beam has been turned off.  This radioactivity can pose a hazard to personnel, who may have to enter the enclosing buildings periodically for maintenance such as target replacement.  Thus, quantifying the severity of this hazard is necessary for compliance with radiological standards (see, for example,~\cite{safety}).

Despite this importance, general procedures for the calculation of this dose are not common.  The typical method of its calculation involves simulating high-energy particle collisions with nuclei and low-energy neutron capture in a Monte Carlo particle simulation program (e.g.~MARS15~\cite{mars}, FLUKA~\cite{fluka-1, fluka-2}, MCNP6~\cite{mcnp6}), producing inventories of residual nuclei, and converting activities of those nuclei after a certain period of time (calculated in MARS15 using DeTra~\cite{detra}) to an estimate of residual dose.  Such methods typically have large uncertainties, typically a factor of two or three~\cite{dose-uncertainty}.  Additionally, such methods are only valid for specific geometries, and adjustments -- some of which require the use of symmetries of a particular irradiated object -- must be made to study small objects or to compute doses at a distance \cite{Mokhov, Pronskikh}. %To estimate the residual dose for various complex geometries and irradiation profiles, the code systems (i.e.~DORIAN~\cite{DORIAN}) are used with other codes (i.e.~the FLUKA code~\cite{fluka-benchmark}).

Instead of these dosimetric methods, it is possible to estimate radiation doses by Monte Carlo simulation of gamma rays emitted from activated materials.  Such an approach is more accurate and more general, albeit more computationally intensive, than dosimetric methods.  FLUKA~\cite{fluka-1, fluka-2} and the FLUKA-based code DORIAN \cite{DORIAN} are two such implementations.  FermiCORD also uses this approach but is based on the MARS15 code and employs the Delauney triangulation for complex geometries not implemented elsewhere. The codes are sequentially used for 1) sectioning the input geometry into parts of appropriate size (involving the Delaunay triangulation), 2) preparing the list of relevant residual nuclides and their $\gamma$-quanta, 3) analysis of the histograms of residual dose on contact, and 4) preparing the $\gamma$-ray sampling routines.

\section{Description of algorithm}
 The algorithm implemented in FermiCORD splits the procedure for calculating residual doses into two stages: a determination of radionuclides and a simulation of the decay of these nuclides.  Both stages of this algorithm rely on the Monte Carlo particle transport code MARS15, a manual for which can be found at \cite{mars}. A flowchart of the algorithm is as follows:
\\
\tikzstyle{steps} = [rectangle, rounded corners, minimum width=3cm, minimum height=1cm, text centered, draw=black, fill=blue!30]
\tikzstyle{stage} = [draw, ellipse, fill=yellow!70, text centered]
\tikzstyle{arrow} = [thick,->,>=stealth]

\begin{tikzpicture}[node distance=1.5cm]
\node (stage1) [stage] {\huge Stage 1};
\node (split) [steps, below of=stage1, align=center, yshift=-0.5cm] {Split regions in\\
the geometry\\
as necessary};
\node (hists) [steps, below of=split, align=center, yshift=-1.5cm] {Set up contact\\
dose histograms in\\
the desired regions};
\node (mars1) [steps, below of=hists, align=center, yshift=-1.5cm] {Run MARS to obtain\\
nuclide inventories and\\
histograms of contact\\
doses in all the regions};
\node (stage2) [stage, right of=stage1, xshift=5.5cm] {\huge Stage 2};
\node (detra) [steps, below of=stage2, align=center, yshift=-0.5cm] {Run DeTra on these\\
nuclide inventories to\\
account for radioactive\\
decay};
\node (filter) [steps, below of=detra, align=center, yshift=-0.7cm] {Filter DeTra's output to\\
preserve only nuclides\\
that can decay by gamma\\
emission with $E > 100$ keV};
\node (routines) [steps, below of=filter, align=center, yshift=-0.5cm] {Write gamma ray\\
sampling subroutines\\
for MARS};
\node (mars2) [steps, below of=routines, align=center, yshift=-0.5cm] {Run MARS in gamma ray\\
sampling mode to generate a\\
histogram of residual doses};

\draw [arrow] (split) -- (hists);
\draw [arrow] (hists) -- (mars1);
\draw [arrow] (detra) -- (filter);
\draw [arrow] (filter) -- (routines);
\draw [arrow] (routines) -- (mars2);
\end{tikzpicture}

\subsection{Preparation of geometry files}
In order to calculate the nuclide distributions accurately, the geometry description file typically requires a few modifications.  First, large regions must be subdivided into smaller ones if large variations in neutron bombardment (and thus in radionuclide production) are present across this material.  Second, each region in which nuclide production will be calculated must be assigned to a unique material.  In general, this requires creating a copy of the relevant material and assigning the region to the copy of the material.

%In the process of subdivision and assignment to new materials, however, the user should be aware that in the MARS15 version used, nuclide distribution can only be calculated in forty materials (and thus forty regions) at a time.  It is recommended in general to limit oneself to no more than forty regions, but if more are necessary, the user must make multiple runs of MARS15 with different region lists.

Many regions must be subdivided manually, but for regions that can be modeled as a prism with polygonal base -- for example, the ceiling of a room -- it often suffices to triangulate the base and then divide the region into triangular prisms.  Since this is often the case, an algorithm was developed to triangulate these regions.  For this purpose, the triangles used should not be too thin and should be as close as possible to equilateral.  A thin triangle will be longer in some direction, and if the radiation exposure varies significantly along the length of the triangle, this will introduce inaccuracies.

In the first step of the algorithm, the algorithm randomly places points inside each of the regions that will form the vertices of the new triangles.  Vertices too close to the boundary of the region, which would produce thin triangles, are rejected.

\begin{figure}
\begin{center}
\begin{subfigure}{0.45\textwidth}
\includegraphics[width=\textwidth]{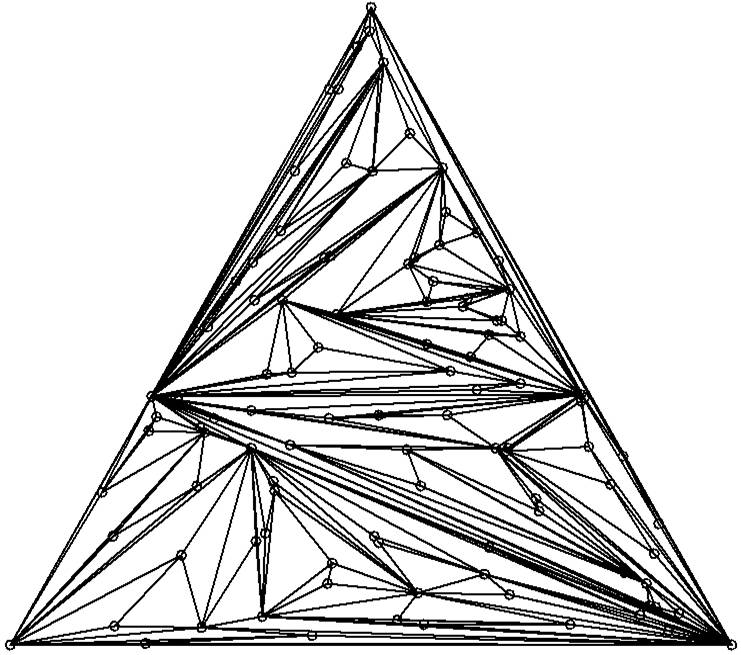}
\caption{}
\label{bad}
\end{subfigure}
\begin{subfigure}{0.45\textwidth}
\includegraphics[width=\textwidth]{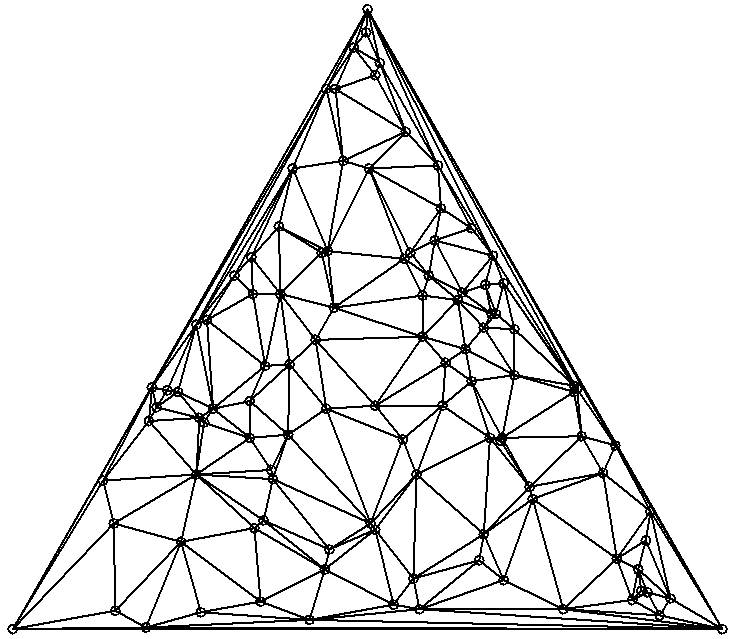}
\caption{}
\label{good}
\end{subfigure}
\caption{An arbitrary triangulation (\subref{bad}) and a Delaunay triangulation (\subref{good}).  The arbitrary triangulation generally includes many long, thin triangles, which are undesirable here, whereas the Delaunay triangulation seeks to avoid these and thus provides adequate resolution in all directions.}
\label{delaunay-figure}
\end{center}
\end{figure}

In the second stage, the added vertices and those on the boundary are triangulated in what is called a Delaunay triangulation.  This triangulation has the properties that the circumcircle of any three points does not enclose any additional points and that the smallest angle in the triangulation is as large as possible.  These properties have the consequence that long, thin triangles (which have large circumcircles and at least one small angle) are avoided (see Figure \ref{delaunay-figure}).  A simple algorithm for accomplishing this is to construct an arbitrary triangulation and then adjust it until it becomes Delaunay.

Following the approach in \cite{Delaunay}, a pair of adjacent triangles that share a common edge is inspected.  If the two angles in the triangles opposite from the common edge have a combined measure of greater than 180 degrees, the smallest angle measure in this pair can be reduced by removing the common edge and replacing it with the edge connecting the other two (previously disconnected) vertices.  This procedure is performed on all pairs of adjacent triangles in the triangulation, but since each edge-flipping step creates new triangles, it is necessary to sweep through the triangulation again and repeat the procedure until it is possible to sweep through the entire triangulation without flipping any edges.\footnote{The triangulation code and Figure \ref{delaunay-figure} were produced in part under an earlier project supported by the National Science Foundation under agreement No. DMS-1055897.}

The code, in its current stage, requires that any regions passed to it be convex.  (A convex shape is one whose interior angles are all less than 180 degrees.)  Thus, the user is manually required to divide the region into convex subregions, which this algorithm will then subdivide into triangles.

\subsection{Stage 1. Production of Radionuclides}
During a simulation of particle transport, MARS15 has the capability to calculate an inventory of radionuclides produced within a material based on collisions of particles with nuclei of that material.  Such inventories are saved for every material that the user specifies, up to a total of 500 materials, corresponding to the upper limit in MARS15~\cite{mars}.  Since these inventories are summed over the entire region corresponding to a material, this has the potential to introduce significant uncertainties for large regions, which motivated the subdivision of these regions into smaller ones.

Even with smaller regions, however, there is some variation in nuclide production within a region.  To estimate this distribution of nuclide production, histograms of residual dose on contact within that region were constructed.  While MARS15's estimation of residual dose on contact is approximate, it should capture the \emph{relative} levels of activation within a material.  The probability density of producing a radionuclide at a point within a region is assumed to be proportional to the residual dose on contact at that point.

The problem of sampling position is difficult in general, especially for irregularly shaped regions (as are often encountered in practice).  An ideal solution would be to construct a three-dimensional histogram of residual dose. In MARS15 that is done via a set of 2D histograms for certain slices, which can be computationally intensive.  Thus, procedures were developed for estimating the distribution from a more limited set of histograms.

\begin{center}
\begin{figure}
\includegraphics[trim = {5cm 4.85cm 16.5cm 5cm}, clip, width=\textwidth]{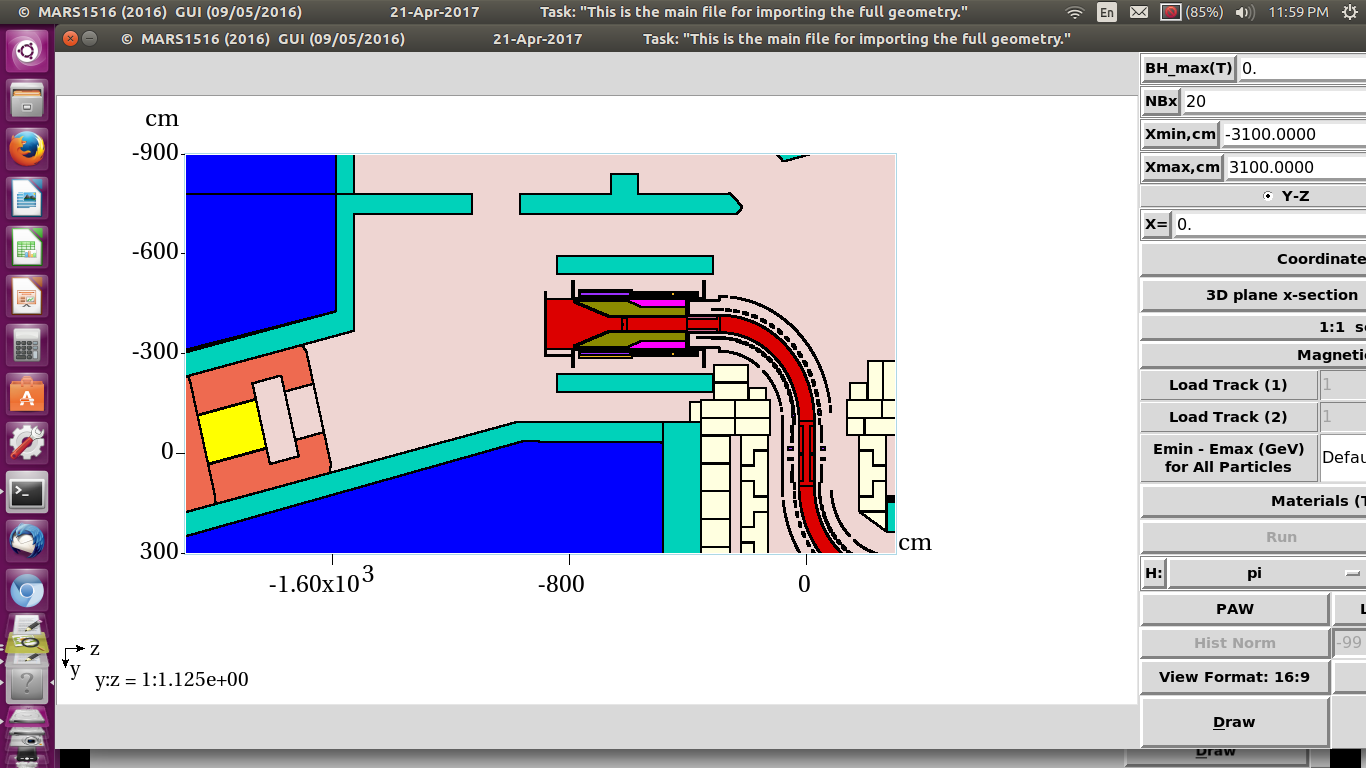}
\caption{A MARS15 model of the Mu2e Production Solenoid Hall.  The target is in the center, and the beam dump is in the lower left-hand corner.  Both axes are in cm.}
\label{mu2e}
\end{figure}
\end{center}

\begin{center}
\begin{figure}
\includegraphics[trim = {1.8cm 3.85cm 12.1cm 3.5cm}, clip, width=\textwidth]{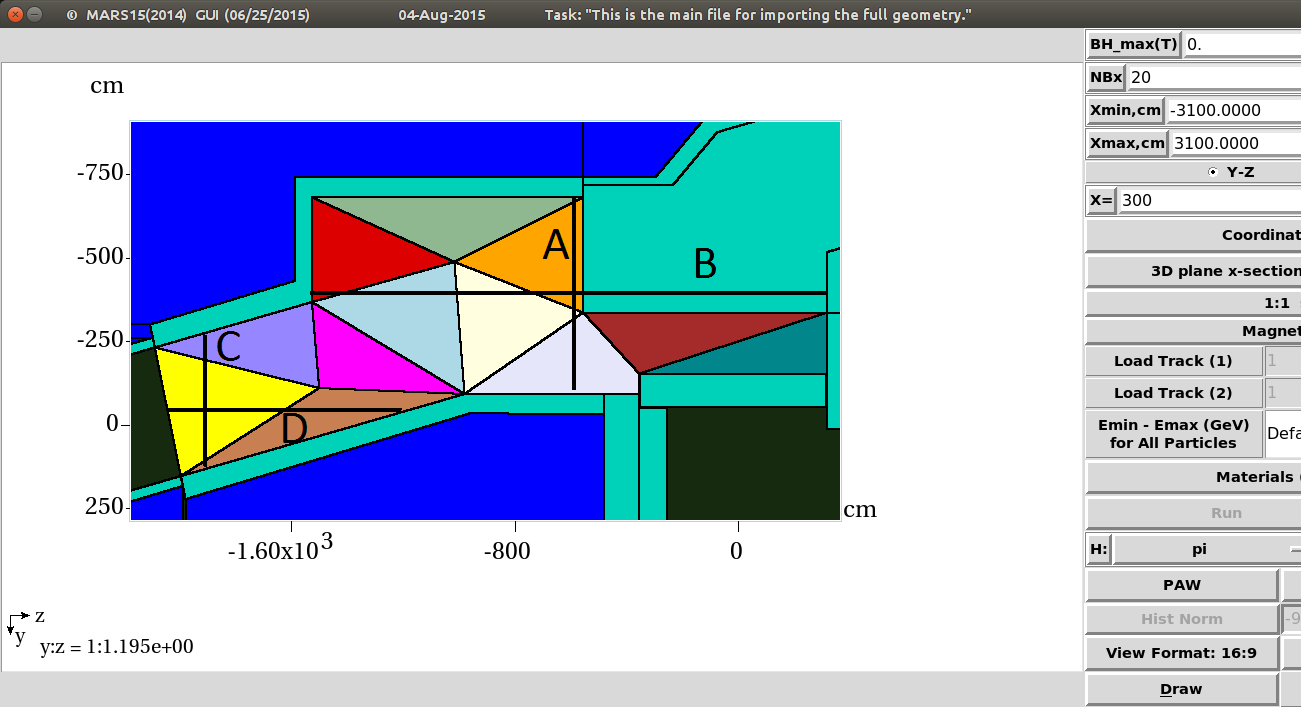}
\caption{A diagram of the ceiling of the Mu2e Production Solenoid Hall.  The division of the ceiling into triangular regions is shown, as are the locations of histograms where residual dose was measured.  The target is located directly below the intersection of histograms A and B; the beam dump is located below the lower left corner.  Both axes are in cm.}
\label{histograms-ceiling}
\end{figure}
\end{center}

\paragraph{Sampling position in ceiling}
As one application of this code, the Mu2e Target Station was studied (see Figure~\ref{mu2e}).  Since the ceiling of this room is irregularly shaped in the horizontal directions but has the same height at all points in consideration, the ceiling was divided into triangles to account for horizontal variation, and histograms were used to estimate vertical variation in nuclide production.

\begin{center}
\begin{figure}
\includegraphics[width=\textwidth]{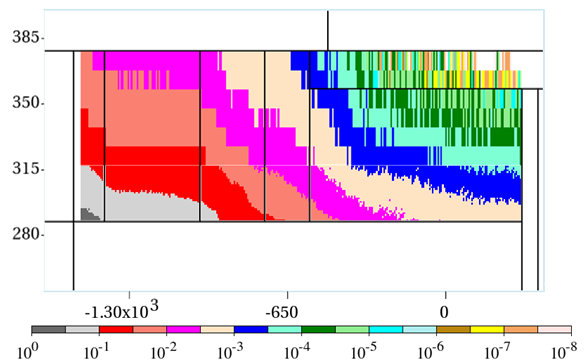}
\caption{A histogram of the depth profile in the ceiling of the Production Solenoid Hall, corresponding to histogram B in Figure \ref{histograms-ceiling}.  The target is located roughly below the midpoint of the histogram; the left half corresponds to the region downstream of the target.  Note that finer bins were used in the lower section of the ceiling, where residual activity is concentrated.  Dimensions of both axes are in cm; residual doses in the ceiling are in mSv/hr, but the exact numerical values are not important for the results discussed here.}
\label{depth-profile}
\end{figure}
\end{center}

Pairs of orthogonal histograms were used to determine the depth profile at various locations in the ceiling.  Figure \ref{histograms-ceiling} illustrates the horizontal position of the histograms; each histogram also extends vertically into the depth of the ceiling.  To reduce computational time, two histograms were used above each of lines A, B, C, and D in Figure \ref{histograms-ceiling}: one with finely divided bins for the lower 30 cm of the ceiling, and one with coarser bins for the remainder of the ceiling, where nuclide production is lower and gammas emitted are more heavily shielded.  The depth profile for the histogram above the target (histogram B) is shown in Figure \ref{depth-profile}.  Histograms C and D are for sampling from the sections of the ceiling near the beam dump (the four triangular regions in the lower left corner); A and B are for the remainder of the ceiling.

Given the horizontal coordinates of a point on the ceiling (which are chosen from a uniform distribution on the triangular region being sampled), this point is projected onto the north-south and east-west histograms.  The depth profiles corresponding to the projections on each of the histograms were averaged to estimate the depth spectrum at the point of interest.  (In the case that the relevant coordinate is outside the range of the histogram, the highest or lowest coordinate value of the histogram was used to obtain the profile.)  The depth into the ceiling was sampled from this spectrum.

While this method was originally written with the ceiling in mind, it can also be used to sample within the floor and walls.  The penetration depth of radiation into the ceiling should be roughly the same as the penetration depth into the walls, so the ceiling histogram can also be used for sampling radiation emitted from the walls and floor.  For determining the depth in regions with irregular orientation, the vector normal to the interior surface must be specified manually. It is calculated and given to the user if one uses the ROOT geometry mode in MARS15.

\paragraph{Other sampling techniques}
The method that was used for the ceiling does not generalize to all regions.  In particular, creating two-dimensional histograms assumed that the region was aligned with the coordinate axes and that the region had a relatively simple geometry.  In practice, these assumptions may not hold for all regions of interest.

For small regions, such as the target, it often suffices to sample uniformly from the region.  For larger regions with significant variation in activity and where there are no symmetry considerations that might permit a simplification (such as an irregularly shaped concrete region that surrounds the beam dump, shown in the lower left corner of Figure \ref{our-dose}), it is possible to approximate a three-dimensional histogram by dividing the region into slices and taking a two-dimensional histogram of each slice.  

\subsection{Stage 2. Decay of Radionuclides}
\label{formulas}
The quantity of various isotopes in a decay chain as a function of time is given by the solution to the Bateman equations, a system of coupled differential equations for radionuclide decay.  These equations can be extended to include production of nuclides from external sources, such as an accelerator beam or a reactor.  The generalized equations are solved using the program DeTra \cite{detra}, which can be called from MARS15.

From DeTra, a list of isotopes with their corresponding concentrations and activities is obtained for each region.  This list of isotopes is compared to the library of gamma rays developed for the SHAMAN nuclear identification system \cite{shaman} to determine the rate of gamma ray production at various energies.  Due to the very short penetration depths of gamma rays with energies less than 100 keV, these are neglected from the analysis (although this threshold can be modified by users). The rate of gamma ray production $r$ of an isotope $I$ is the product of the activity $A$ (the number of nuclear decays per second) and the number of gamma rays per decay above the 100 keV threshold, which is the sum of the branching ratios $p_j$ corresponding to these energies:
$$ r_I = A_I\sum_{\mathclap{j:\ E_j > 100\ \text{keV}}} p_j $$
Positrons emitted are assumed to annihilate immediately to produce two 511 keV gamma rays.

When sampling gamma emission, a region is randomly selected based on its total rate of gamma production, and a gamma energy is randomly selected based on the relative production rates of gammas within that region.  The point of emission within the region is chosen based on the histograms of residual dose (as described above), and the angle of emission is assumed to be isotropic.  This gamma ray is then tracked using MARS15, with corresponding energy deposition at relevant points computed using a separate histogram in the interior of the room.

\section{Validation of FermiCORD}
To determine the validity of FermiCORD, the results were compared to another Monte Carlo simulator, the FLUKA-based code DORIAN, and also to experimental results directly from the CERN-EU High-Energy Reference Field Facility (CERF).

\subsection{Comparison to DORIAN}
Since FLUKA has been benchmarked to measurements at accelerator facilities and cosmic ray experiments \cite{fluka-benchmark}, comparison to a FLUKA-based simulation should give some indication of FermiCORD's accuracy.  In \cite{DORIAN}, a simple simulation in DORIAN is described in which a 0.433 GeV proton beam impinges on a copper target with radius 5 cm and length 50 cm.  The target is irradiated for one year and then allowed to cool for a variable length of time before residual dose is measured 50 cm upstream of the target.

This simulation was repeated in FermiCORD.  The target was subdivided into forty 1.25 cm-thick slices to account for variations in nuclide production along the length of the target.  Since the distance between the point of measurement and the target is much larger than the target radius, the radial distribution of nuclide production in the target does not significantly affect the calculated dose and was therefore ignored here.  As a result, division of the target into regions was sufficient for estimating the distribution; no histograms were needed.

%Whereas DORIAN provides the residual dose as a continuous function of time (calculated from $\sim 10^3$ sec to $\sim 10^8$ sec in \cite{DORIAN}), FermiCORD calculates the residual dose after a pre-determined cooling time.  For comparison purposes, cooling times of $10^3$ seconds, $10^4$ seconds, 1 day $\sim 10^5$ seconds, 12 days $\sim 10^6$ seconds, 116 days $\sim 10^7$ seconds, and 1157 days $\sim 10^8$ seconds were investigated.

The results are compared to DORIAN in Table \ref{DORIAN-table}.  % Figure \ref{DORIAN-target}.
In this simulation, FermiCORD's results agree with those from DORIAN to within about 10\% or better at all cooling times investigated.

\begin{table}
\begin{center}
\begin{tabular}{|c|c|c|}
\hline
Cooling time & Residual Dose (FermiCORD) & Residual Dose (DORIAN) \\ \hline
$10^3$ sec & $2.3 \times 10^{-8}$ & $2.5 \times 10^{-8}$ \\ \hline
$10^4$ sec & $1.8 \times 10^{-8}$ & $1.8 \times 10^{-8}$ \\ \hline
1 day & $1.5 \times 10^{-8}$ & $1.4 \times 10^{-8}$ \\ \hline
12 days & $1.0 \times 10^{-8}$ & $9.9 \times 10^{-9}$ \\ \hline
116 days & $3.8 \times 10^{-9}$ & $3.7 \times 10^{-9}$ \\ \hline
1157 days & $3.1 \times 10^{-10}$ & $2.8 \times 10^{-10}$ \\ \hline
\end{tabular}
\caption{The residual dose from a Cu target after 1 year of irradiation and various cooling times calculated by FermiCORD and DORIAN.  Residual doses are in units of $\mu$Sv/hr and are normalized to a beam intensity of 1 p/s.  The values for the DORIAN code are taken from Figure 2 of \cite{DORIAN}.}
\label{DORIAN-table}
\end{center}
\end{table}

\subsection{Comparison to CERF data}
A more direct check to FermiCORD is to compare the results to experimental data.  At the CERF facility at CERN, a beam of 120 GeV/$c$ protons and pions is incident upon a copper target, and samples of various materials (aluminum, concrete, copper, iron, titanium, and others not considered here) are placed either downstream or to the side of the target.  The samples were irradiated with on the order of $10^{12}$ particles over a period of several days (the exact number varied from sample to sample) and then allowed to cool.  Measurements of the dose emitted by the sample were taken 12.4 cm away from the sample at cooling times ranging from about an hour to several days \cite{cerf-1, cerf-2}.  More detailed parameters of the experimental setup are contained in the SINBAD database \cite{sinbad}.

FermiCORD demonstrated good agreement with the CERF data for iron and copper.  The titanium sample showed good agreement at short cooling times but worse agreement at longer times, and the concrete and aluminum samples had consistently lower doses in FermiCORD than what was measured experimentally.  For all samples considered, however, the disagreement was never worse than a factor of about 2.

The results from FermiCORD and CERF are displayed below in Tables \ref{cerf-table-al} through \ref{cerf-table-ti}.

% Add to all tables a last column -- calculation to experiment ratio (name it C/E).

\begin{table}
\begin{center}
Aluminum
\begin{tabular}{|c|c|c|c|}
\hline
Cooling time & Calculated Dose & Experimental Dose & Calculation / Experiment \\ \hline
1.17 & 41.5 $\pm$ 0.5 & 83 $\pm$ 4 & 0.50 \\ \hline
3.12 & 34.3 $\pm$ 0.7 & 74 $\pm$ 4 & 0.46 \\ \hline
16.95 & 15.0 $\pm$ 0.3 & 39 $\pm$ 2 & 0.38 \\ \hline
24.50 & 10.8 $\pm$ 0.3 & 26 $\pm$ 2 & 0.41 \\ \hline
\end{tabular}
\caption{The irradiation profile of an aluminum sample following irradiation in the CERF facility at CERN, obtained from simulation in FermiCORD and directly measured experimentally.  Cooling times are in hours and residual doses are in nSv/hr.  The experimental values for the CERF facility are shown in Figure 1 of \cite{cerf-2}.
\\\hspace{\textwidth}
Uncertainties for the calculated doses reflect statistical uncertainties alone (and ignore, among other effects, uncertainties in the cross-sections for radionuclide production used as input to MARS15, which can be 50\% or larger).  The uncertainties in the experimental data reflect the precision of the detector ($\pm 1$ nSv/hr) and a 2.3\% uncertainty in the source-to-detector distance (which translates to a 4.6\% uncertainty in dose).}
\label{cerf-table-al}
\vspace{10pt}
Concrete
\begin{tabular}{|c|c|c|c|}
\hline
Cooling time & Calculated Dose & Experimental Dose & Calculation / Experiment \\ \hline
1.17 & 20.1 $\pm$ 1.9 & 42 $\pm$ 2 & 0.48 \\ \hline
1.93 & 11.5 $\pm$ 1.6 & 22 $\pm$ 1 & 0.52 \\ \hline
6.48 & 5.2 $\pm$ 0.8 & 10 $\pm$ 1 & 0.52 \\ \hline
\end{tabular}
\caption{Same as Table \ref{cerf-table-al} but for concrete.  The experimental values are shown in Figure 4 of \cite{cerf-2}.}
\label{cerf-table-concrete}
\vspace{10pt}
Copper
\begin{tabular}{|c|c|c|c|}
\hline
Cooling time & Calculated Dose & Experimental Dose & Calculation / Experiment \\ \hline
1.27 & 75.4 $\pm$ 4.4 & 61 $\pm$ 3 & 1.24 \\ \hline
2.02 & 65.2 $\pm$ 3.3 & 55 $\pm$ 3 & 1.19 \\ \hline
12.58 & 25.0 $\pm$ 1.5 & 26 $\pm$ 2 & 0.96 \\ \hline
43.88 & 11.0 $\pm$ 0.3 & 12 $\pm$ 1 & 0.92 \\ \hline
\end{tabular}
\caption{Same as Table \ref{cerf-table-al} but for copper.  The experimental values are shown in Figure 2 of \cite{cerf-2}.}
\label{cerf-table-cu}
\vspace{10pt}
Iron
\begin{tabular}{|c|c|c|c|}
\hline
Cooling time & Calculated Dose & Experimental Dose & Calculation / Experiment \\ \hline
1.17 & 137.0 $\pm$ 2.6 & 149 $\pm$ 7 & 0.92 \\ \hline
2.85 & 111.0 $\pm$ 2.8 & 117 $\pm$ 5 & 0.95 \\ \hline
18.32 & 70.1 $\pm$ 1.7 & 70 $\pm$ 3 & 1.00 \\ \hline
67.77 & 51.7 $\pm$ 0.9 & 43 $\pm$ 2 & 1.20 \\ \hline
\end{tabular}
\caption{Same as Table \ref{cerf-table-al} but for iron.  The experimental values are shown in Figure 3 of \cite{cerf-2}.}
\label{cerf-table-fe}
\vspace{10pt}
Titanium
\begin{tabular}{|c|c|c|c|}
\hline
Cooling time & Calculated Dose & Experimental Dose & Calculation / Experiment \\ \hline
1.13 & 148.6 $\pm$ 1.3 & 156 $\pm$ 7 & 0.95 \\ \hline
2.22 & 124.3 $\pm$ 3.1 & 133 $\pm$ 6 & 0.93 \\ \hline
14.10 & 35.9 $\pm$ 0.8 & 59 $\pm$ 3 & 0.61 \\ \hline
56.77 & 13.7 $\pm$ 0.6 & 24 $\pm$ 1 & 0.57 \\ \hline
\end{tabular}
\caption{Same as Table \ref{cerf-table-al} but for titanium.  The experimental values are shown were obtained from the SINBAD database \cite{sinbad}.}
\label{cerf-table-ti}
\vspace{10pt}
\end{center}
\end{table}

\section{Calculation of the residual dose for the Mu2e Target Station}
The future Mu2e experiment at Fermilab~\cite{mu2e} will attempt to observe the neutrinoless conversion of muons into electrons, which, if discovered, would reveal physics beyond the Standard Model.  To generate the muons, a high-intensity proton beam ($6\times 10^{12}$ protons per second with energy 8 GeV) will impinge on a tungsten target, producing pions that will then decay into muons.  A beam dump is located behind the target to capture particles produced in the collisions, although many particles instead strike the walls, ceiling, floor, and other structures in the room.

The descriptions of the geometry and of the magnetic fields were taken from a proposed design for the experiment~\cite{mu2e}.  The irradiation and cooling times were chosen to be 1 year and 1 week, respectively, as the experiment is expected to shut down annually for maintenance.
% Note - how can we better describe the framework model?

In this simulation emissions from the target, beam dump and surrounding concrete, heat and radiation shield, end cap, walls, ceiling, and floor were considered.  The ceiling and floor were triangulated as described above, and the walls were divided into sections as well.

Residual radiation levels in the Production Solenoid hall, where the target and beam dump are located, have previously been estimated.  One such calculation was made in \cite{Leveling}, which considers only the activity of the region surrounding the target and the beam dump (and thus excludes contributions from the walls, floor and ceiling).  Additionally, the sampling methods used in \cite{Leveling} were less sophisticated than those in FermiCORD: the target was assumed to be a point source, the heat and radiation shield and the end cap were sampled uniformly.  The beam dump was divided into vertical slices parallel to the front face, but within each slice, all the activity was assumed to emanate from a cylindrical region near the center of the slice.

\begin{figure}[!htb]
\begin{center}
\begin{subfigure}{0.45\textwidth}
\includegraphics[width=\textwidth]{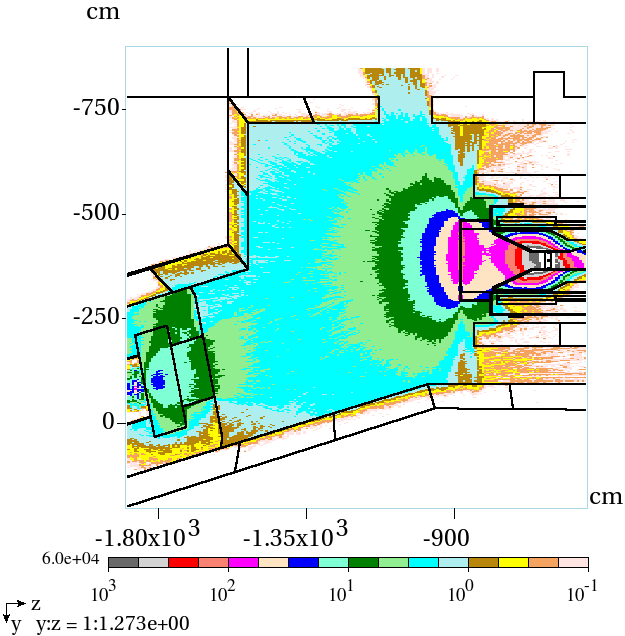}
\caption{}
\label{our-dose}
\end{subfigure}
\begin{subfigure}{0.45\textwidth}
\includegraphics[width=\textwidth]{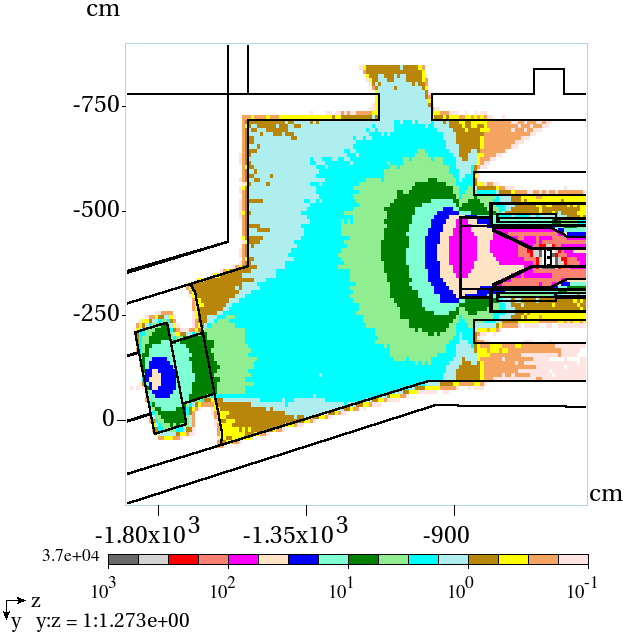}
\caption{}
\label{leveling-dose}
\end{subfigure}
\end{center}
\caption{A comparison of the residual dose calculated in the Mu2e Production Solenoid hall using FermiCORD (\subref{our-dose}) and an approximate method \cite{Leveling} based on sampling photons from few sources (\subref{leveling-dose}).  Doses assume one year of irradiation and one week of cooling and are in units of mSv/hr.}
\label{PShall-graph}
\end{figure}

The original calculation in \cite{Leveling} was performed using a different proposal for the Mu2e design, but, for comparison purposes, the calculations were modified to match the design that was examined with FermiCORD.\footnote{In order to apply the method in \cite{Leveling} to the geometry used in FermiCORD, the method had to be re-implemented, and during the re-implementation process, some changes were made to the method.  In particular, the original method also considered radiation from additional sources of radiation in the vicinity of the target, such as the metal spokes holding the target in place.}  A comparison between FermiCORD and this revised calculation is shown in Figure \ref{PShall-graph}.  In particular, the dose calculated in corners is noticeably higher using FermiCORD due to the radiation emitted by the walls.

%FermiCORD predicted doses about half the level of the previous estimate.  However, in addition to the differences between the sampling methods used in these estimates and the larger number of sources used in FermiCORD, the estimates were also made using different models of the proposed geometry~(\cite{nd2013, review}), and all of these factors affect the calculated dose distribution.

\begin{figure}
\begin{center}
\begin{subfigure}{0.7\textwidth}
\includegraphics[trim={2.1cm 2.7cm 13.4cm 3.7cm}, clip, width=\textwidth]{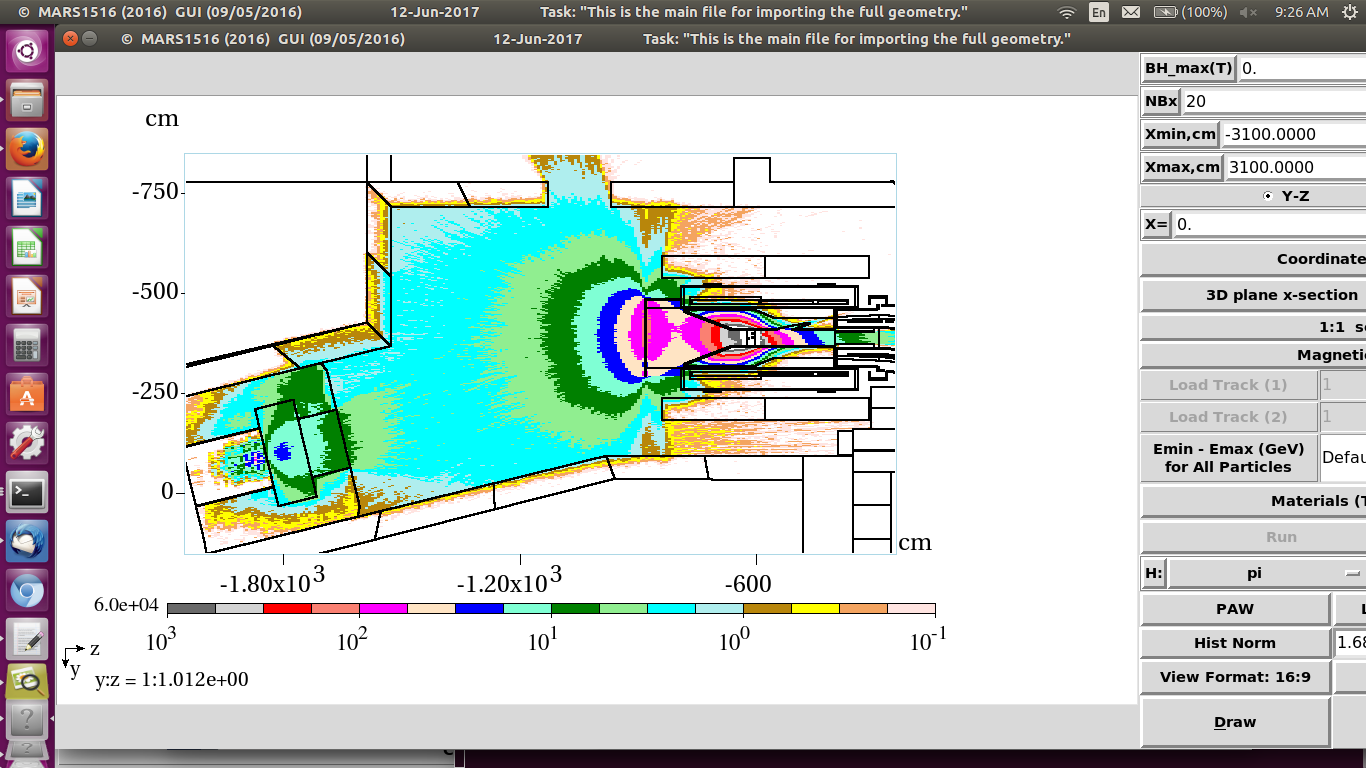}
\caption{}
\label{target-plane}
\end{subfigure}
\begin{subfigure}{0.7\textwidth}
\includegraphics[trim={2.1cm 2.7cm 13.4cm 3.7cm}, clip, width=\textwidth]{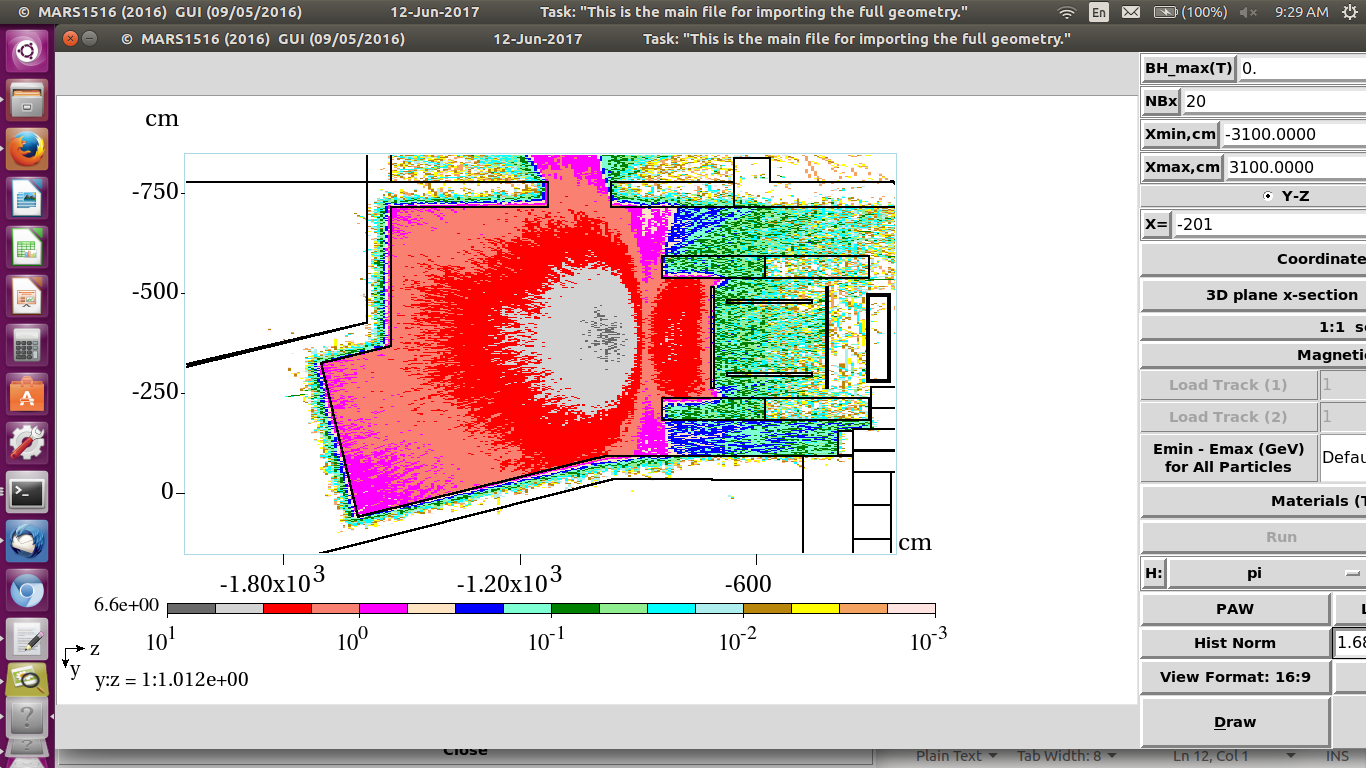}
\caption{}
\label{floor-plane}
\end{subfigure}
\begin{subfigure}{0.7\textwidth}
\includegraphics[trim={2.1cm 2.7cm 13.4cm 3.7cm}, clip, width=\textwidth]{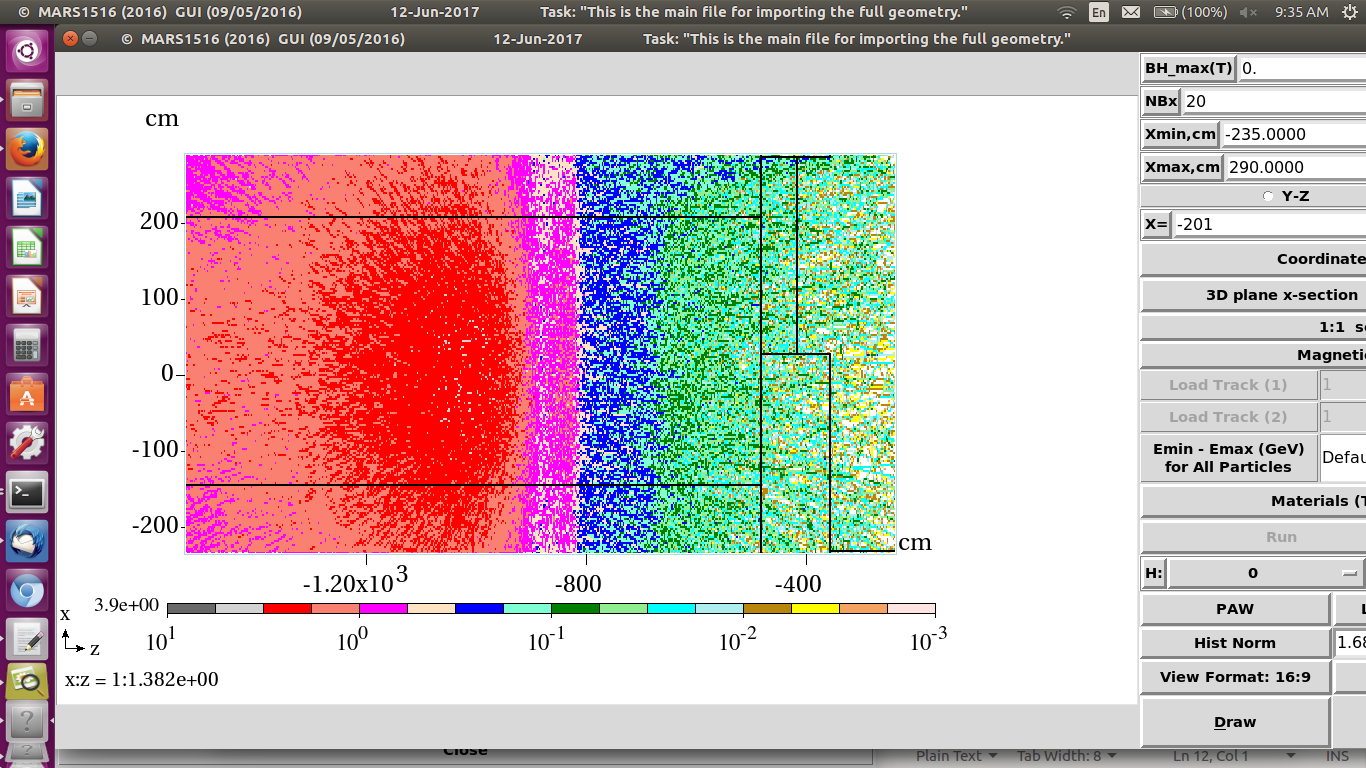}
\caption{}
\label{north-wall}
\end{subfigure}
\caption{Plots of residual doses (in mSv/hr) in the Mu2e Production Solenoid hall in the horizontal plane of the target (\subref{target-plane}), in the horizontal plane $\sim 30$ cm above the floor (\subref{floor-plane}), and in the vertical plane $\sim 40$ cm away from the north wall (\subref{north-wall}).   Doses assume one year of irradiation and one week of cooling.}
\label{slices}
\end{center}
\end{figure}

In Figure \ref{slices}, residual doses calculated with FermiCORD at various positions in the Mu2e Production Solenoid hall are displayed.

%Since FermiCORD calculates contributions to the dose not only from the target and beam dump but also from the walls, ceiling, and floor, estimates of doses near the walls (and away from the target) should be reliable.  In Figure \ref{slices}, residual doses calculated at various positions in the Production Solenoid hall are displayed.

% Are there any results from Leveling in the comparable positions that we can compare to?  I think there were some pictures of Leveling's results, but I don't think the locations exactly correspond.

\section{Conclusion}
The code system FermiCORD for the MARS15 code for calculation of the accelerator-induced residual dose in complex geometries and for arbitrary irradiation profiles has been developed, benchmarked, and applied to simulations of the residual dose in the Mu2e Production Solenoid Hall. The code system works with the MARS15 code and requires two stages: one for calculation of the inventories of residual nuclides in the components of the facility of given topology and one for sampling the emission of secondary photons and scoring the dose at the locations of interest. Calculations indicate that although the approximate approach consisting in scoring the residual dose from only the few most radioactive components is rather simple and less time consuming, an accurate determination of the dose at the remote locations of particular importance for the safety of personnel requires a simulation relying on the full set of sources of radioactivity. The FermiCORD system for the MARS15 code accounts for the full set of sources and therefore should be more accurate.

\section{Acknowledgements}
We are grateful for the vital contributions of the Fermilab staff and the technical staff of the participating institutions.  This work was supported by the US Department of Energy; the Italian Istituto Nazionale di Fisica Nucleare; the Science and Technology Facilities Council, UK; the Ministry of Education and Science of the Russian Federation; the US National Science Foundation; the Thousand Talents Plan of China; the Helmholtz Association of Germany; and the EU Horizon 2020 Research and Innovation Program under the Marie Sklodowska-Curie Grant Agreement No.690385. Fermilab is operated by Fermi Research Alliance, LLC under Contract No. De-AC02-07CH11359 with the US Department of Energy, Office of Science, Office of High Energy Physics. The United States Government retains and the publisher, by accepting the article for publication, acknowledges that the United States Government retains a non-exclusive, paid-up, irrevocable, world-wide license to publish or reproduce the published form of this manuscript, or allow others to do so, for United States Government purposes.  This work was supported in part by the U.S. Department of Energy, Office of Science, Office of Workforce Development for Teachers and Scientists (WDTS) under the Science Undergraduate Laboratory Internships Program (SULI).

%\section*{References}

\bibliography{paper-nim}

\end{document}